\documentclass[12pt]{iopart}
\usepackage{iopams}  
\usepackage[breaklinks=true,colorlinks=true,linkcolor=blue,urlcolor=blue,citecolor=blue]{hyperref}
\usepackage{rotating}
\usepackage{epsfig}
\usepackage{graphicx}
\usepackage{epstopdf}

\begin{document}

\pagenumbering{arabic}

\title{Vortex-glass transformation within the surface superconducting state of $\beta$-phase Mo$_{1-x}$Re$_x$ alloys}
\author{Shyam Sundar$^{1,2}$, M. K. Chattopadhyay$^{1,2}$, L. S. Sharath Chandra$^{2}$, R. Rawat$^{3}$ and S. B. Roy$^{1,2}$}
\address{$^{1}$Homi Bhabha National Institute, Raja Ramanna Centre for Advanced Technology, Indore, Madhya Pradesh-452013, India}
\address{$^{2}$Magnetic and Superconducting Materials Section, Raja Ramanna Centre for Advanced Technology, Indore, Madhya Pradesh-452013, India}
\address{$^{3}$UGC-DAE Consortium for Scientiﬁc Research, Khandwa Road, Indore, Madhya Pradesh-452001, India}

\vspace{10pt}
\ead{shyam@if.ufrj.br; shyam.phy@gmail.com}

\vspace{10pt}
\begin{indented}
\item[] August 2016
\end{indented}

\begin{abstract}

We have performed an experimental study on the temperature dependence of electrical resistivity $\rho$($T$) and heat capacity $C$($T$) of the Mo$_{1-x}$Re$_x$ $(x = 0.20, 0.25)$ alloy superconductors in different magnetic fields. In the presence of applied magnetic field, the electrical resistivity of these alloys go to zero at a temperature well above the bulk superconducting transition temperature obtained with the help of heat capacity measurements in the same magnetic field. Our study indicates the presence of surface superconducting state in these alloys, where the flux lines are pinned in the surface sheath of the superconductor. The configuration of the flux-lines (2d  pancake-like) in the surface sheath is understood in the realm of the flux-spot model. Experimental evidence in support of the surface mixed-state state or "Kulik vortex-state" and the occurrence of a vortex-liquid to vortex-glass transition is presented. 

\end{abstract}

%\pacs{}
\vspace{2pc}

%\submitto{\SUST}

\maketitle

\section{Introduction}

\begin{figure}[h]
\centering
\includegraphics[height=8cm]{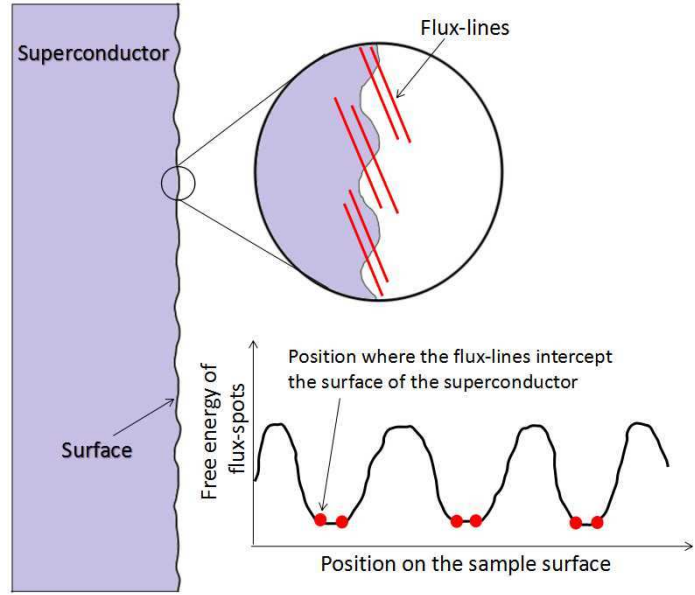}
\caption{\label{fig:fluxspot} Schematic diagram explaining the concept of the flux spot model \cite{HP-1967}.}
\end{figure}

The phenomenon of surface superconductivity was discovered by Saint-James and de Gennes back in 1963 \cite{DP-1963, CY-1964}. They showed that for any finite size sample, the nucleation of the superconducting region is easy to occur near the sample surface in the case of a metal-insulator (or metal-vacuum) interface, in the presence of a parallel magnetic field $H_{c3}$ higher than the upper critical field $H_{c2}$ by a factor of 1.695. They also argued that the nucleation field $H_{c3}$ for surface superconductivity may be strongly modified by the normal metal coating (e.g. Cu, Ag) at the sample surface \cite{GY64, FI66}. With the discovery of surface superconductivity, it was possible to explain a large amount of experimental data, which had previously been discarded  as due to the sample inhomogeneity \cite{Mic96}. Later, many researchers studied the surface superconducting state in many superconductors such as - Pb-Tl, Nb, MgB$_{2}$ \cite{J-1964, HL-1965, LH-1966, Das08, kar70, men06, ryd03}. Initially it was thought that for surface superconductivity (and surface critical current) to exist, the local magnetic field needs to be parallel to the sample surface \cite{HL-1965, abr65, par65, par66, fin65}. Subsequent work, however, showed that surface superconductivity (and surface critical current) may be observed even when the local magnetic field has a nonzero perpendicular component arising because of the misalignment of the applied magnetic field relative to the sample surface, due to the magnetic field related to an applied transport current, the demagnetization factor and the local roughness of the sample surface \cite{HP-1967}. It was inferred \cite{HP-1967} that the perpendicular component of the magnetic field ($H$ $>$ $H_{c2}$) may penetrate the sample surface as an array of quantized flux-spots ($\Phi_0 = h/2e \approx 2.0678 \times10^{-15}$ Wb) whose free energy is sensitive to the local properties of the surface sheath of the superconductor. Since the local properties of the surface sheath will vary spatially, the free energy of the flux spots will also vary spatially and the flux spots will stay pinned at the locations of minimum free energy (see Fig. \ref{fig:fluxspot}) \cite{HP-1967}. The pinning of the flux spots depends on the degree of surface roughness, and becomes increasingly significant till the scale of roughness becomes comparable with the spacing between the flux spots approximately \cite{HP-1967}. This suggests the existence of a dense quantized 2d flux lattice (or flux-spots) in the surface region of superconductor \cite{HP-1967}. The absence of surface pinning or the presence of a  driving force (due to a transport current in the presence of magnetic field) greater than the pinning strength, allows the array of flux-spots to move freely in the surface superconducting region producing a non-zero voltage and a vanishing surface critical current \cite{HP-1967}. It has also been shown that these 2d flux-lines have the same kind of flux-line dynamics as that of  the bulk (3d) superconducting mixed state of the type II superconductors, e.g., the flux flow and the flux creep effects \cite{HP-1967, MRW-1969, JJJ-1970, PBY-1993, AP-2004}, and in the presence of disorder, may enter a critical state, known as the Kulik vortex-state \cite{KUL-1969, KUL-1967}. In-spite of a variety of similar flux-line dynamics as in the bulk mixed state, there is no report of surface vortex-glass, which can act as a 2d counterpart of the vortex-glass state in the superconductors. However, the existence of a true 2d vortex-glass state at any finite temperature is highly controversial. Theoretically, the 2d vortex-glass state is possible only at $T $= 0 K, which has been supported by many authors \cite{MTA-1991, DMD-1991, CPRBA-1992}. Nevertheless, there are some reports, which suggest the existence of 2d vortex-glass even at a finite temperature \cite{MS-1999, B-2000, Y-1992, PBP-2003}. 

Superconductivity in the Mo$_{1-x}$Re$_x$ alloys is quite interesting in the sense that the $T_c$ in these alloys is enhanced by an order of magnitude, as compared to their constituent elements (Mo and Re) \cite{shy14, shy15a}. Apart from the unique superconducting and normal state properties, the Mo$_{1-x}$Re$_x$ alloys are technologically important in terms of the fabrication of superconducting radio frequency (SCRF) cavities \cite{and89}. Another unusual aspect associated with the superconducting state of the Mo$_{1-x}$Re$_x$ alloys is the presence of a surface superconducting mixed state \cite{LER-1966, JOI-64}. In present paper, the temperature dependence of electrical resistivity, $\rho(T)$ across the superconducting transition in various applied magnetic fields has been measured, to explore the surface mixed state of $\beta$-phase Mo$_{1-x}$Re$_x$ $(x = 0.20, 0.25)$ alloys. The bulk superconducting transition temperature has been obtained by measuring the temperature dependence of heat capacity in different applied magnetic fields. It is found that the resistivity goes to zero in the surface superconducting state well above the onset of the bulk superconducting transition, due to the pinning of the flux-lines in the surface superconducting region. This flux-line pinning also leads to a vortex-glass state in the surface superconducting state. The estimated critical exponents corresponding to the vortex-glass to liquid transition is found to be in-line with the 2d nature of the vortex-glass state. A detailed $H$-$T$ phase diagram is presented.

\section{Experimental details}

Polycrystalline samples of Mo$_{1-x}$Re$_x$, ($x$ = 0.20, 0.25) alloys were prepared by melting molybdenum and rhenium (99.95+ \% Purity) in an arc furnace under high purity (99.999 \%) argon atmosphere. The samples were flipped and re-melted six times to ensure the homogeneity. The X-ray diffraction measurement shows that the samples have formed in the single phase bcc crystal structure (the XRD results are presented elsewhere) \cite{shy15b}. The electrical resistivity of the Mo$_{1-x}$Re$_x$ alloy samples were measured in the temperature range 1.5-15 K and up to 2~T magnetic field using the standard four probe technique with the help of a superconducting magnet cryostat (Oxford Instruments, UK) system. The temperature dependence of electrical resistivity in 0, 0.5, and 1 T magnetic fields were also measured after applying a Cu coating on the Mo$_{0.75}$Re$_{0.25}$ alloy and a Ag coating on the Mo$_{0.80}$Re$_{0.20}$ alloy. The Cu coating on the Mo$_{0.75}$Re$_{0.25}$ alloy was applied using the electroplating technique (with the help of a CuSO$_4$ solution). Some portions of the Cu coating was then removed by filing for putting the four probe electrical contacts. The Ag coating on the Mo$_{0.80}$Re$_{0.20}$ alloy was put by dipping the sample in electrically conducting Ag paint after putting the electrical contacts for the four probe technique. A constant current of 100~mA was used to measure the resistivity of these alloys. Heat capacity measurements were performed in the temperature range 2-15~K in various magnetic fields up to 3~T using a Physical Property Measurement System (PPMS, Quantum design, USA).    

\section{Results and Discussion}

	\begin{figure}[h]
	\centering
	\includegraphics[height=9cm]{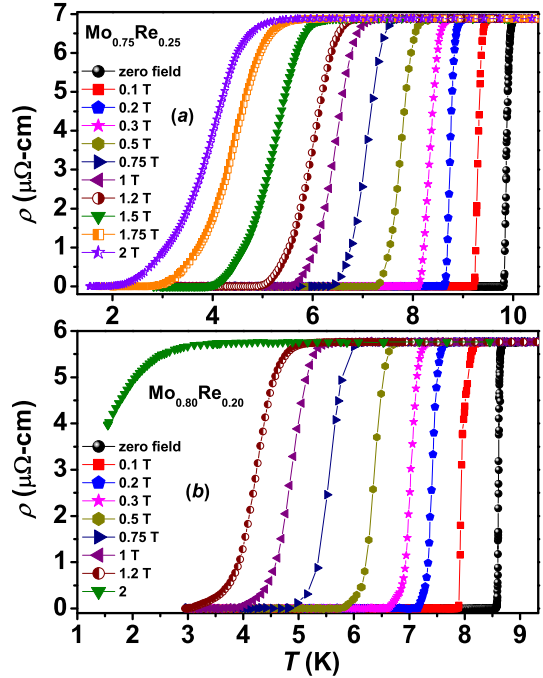}
	\caption{\label{fig:RTMS} Temperature dependence of electrical resistivity, $\rho(T)$ of the Mo$_{1-x}$Re$_x$ alloys measured across the normal to superconducting phase transition in different magnetic fields.}
	\end{figure}
	
Figure \ref{fig:RTMS} shows the temperature dependence of electrical resistivity in zero and different applied magnetic fields. The zero field superconducting transition temperatures ($T_c$) for the present alloys have been reported elsewhere \cite{shy15b} and they are in good agreement with previously reported values \cite{MOR-1963}. The width of the superconducting transition ($\Delta T_c$) for different field values was estimated as the temperature difference between the 10\% and 90\% of the normal state resistivity ($\rho_N$). 

	\begin{figure}[h]
	\centering
	\includegraphics[height=9cm]{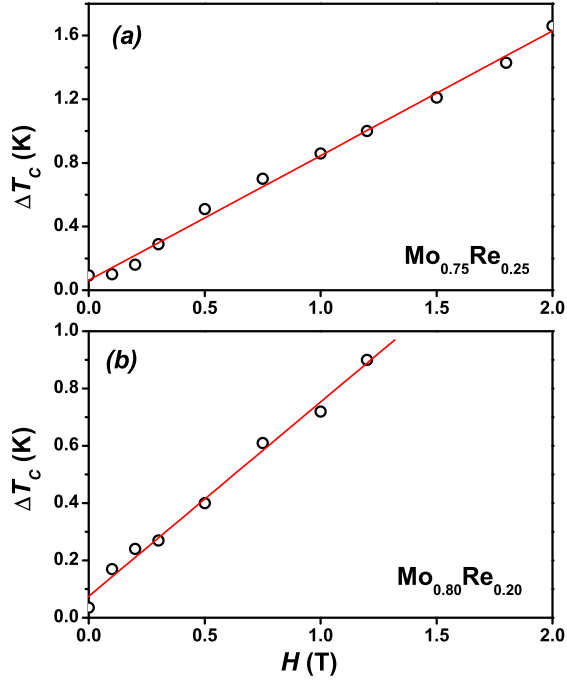}
	\caption{\label{fig:DTCvsH} Magnetic field dependence of the width of the superconducting transition ($\Delta T_c$) in the Mo$_{1-x}$Re$_x$ alloys . The solid lines in the panels $(a)$ and $(b)$ show the fitting of the experimental data using a linear field dependence (Eq. \ref{eq:8-1}).}
	\end{figure}

In Fig. \ref{fig:DTCvsH}, the $\Delta T_c$ as a function of magnetic field is plotted, which is well described using a linear relation (Eq. \ref{eq:8-1}), 

\begin{equation} \label{eq:8-1}
\Delta T_c(H) = \Delta T_c(0) + kH
\end{equation}
where, $\Delta T_c(0)$ is the width of superconducting transition in zero field and $k$ is a constant. The presence of multiple superconducting phases in a sample (and the resulting spatial distribution of $T_c$) has often been considered to be a possible reason behind the broadening of the normal-to-superconducting phase transition \cite{CRXS-2010, JSCM-1998}. The fitting of Eq. \ref{eq:8-1} to the $\Delta T_c$ versus $H$ data, however, negates this possibility. A sample with multiple superconducting phases would have shown an upturn (or curvature) in the low field side of the $\Delta T_c$ versus $H$ plot due to the different field dependencies of $T_c$ in the different superconducting phases \cite{CRXS-2010}. Moreover, the XRD and optical metallography study of both the Mo$_{1-x}$Re$_x$ $(x = 0.20, 0.25)$ alloys \cite{shy15b}, do not indicate the presence of a second phase.

	\begin{figure}[h]
	\centering
	\includegraphics[height=8cm]{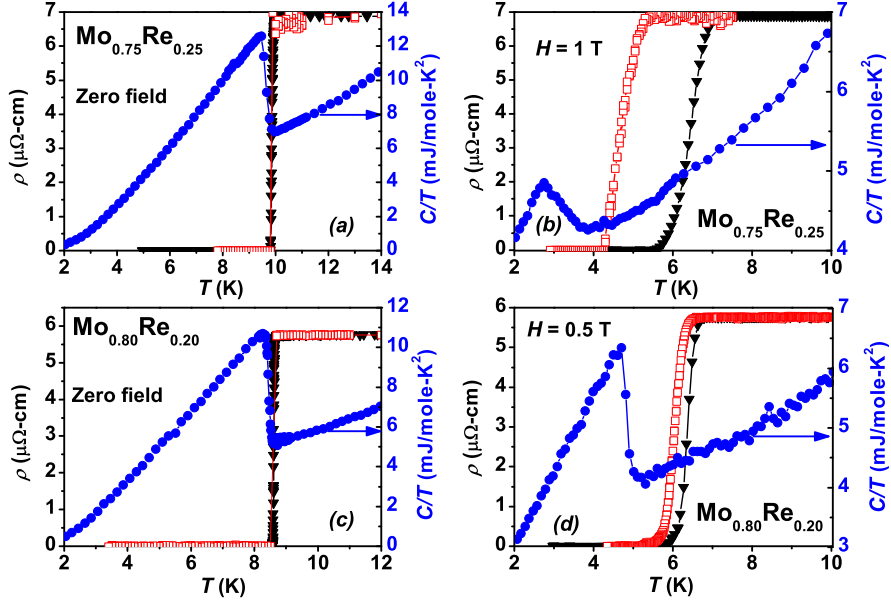}
	\caption{\label{fig:surfacesuperconductivity} \textbf{Panel ($a$) and ($c$):} Temperature dependence of electrical resistivity $\rho(T)$ and heat capacity $C(T)$ in zero magnetic field across the superconducting transition. \textbf{Panel ($b$) and ($d$):} $\rho(T)$ and $C(T)$ in $H$ = 1 T and $H$ = 0.5 T across the superconducting transition for $x$ = 0.25 and $x$ = 0.20 respectively. In all the panels, the (blue) solid circles (dots) represent the $C(T)$ measured in different $H$ (specified in each of the panels), the (black) solid triangles represent the $\rho(T)$ measured before coating the samples with Cu or Ag, and the (red) open squares represent the $\rho(T)$ measured after putting these metal coatings.}
	\end{figure}

Apart from the existence of multiple superconducting phases in the sample, the broadening of the superconducting transition may also be due to the presence of multiple superconducting gaps \cite{AMZYKHHT-2002, Yu-2003}, vortex-glass state \cite{HYYSIGK-2009, BAYOM-1998}, vortex-melting behaviour \cite{Yu-2003} and surface superconductivity \cite{UAGWG-2003, AJSJH-2002}. In this context, surface superconductivity has already been reported in Mo$_{1-x}$Re$_x$ alloys \cite{LER-1966}. On the other hand, the superconducting properties of $\beta$-phase Mo$_{1-x}$Re$_x$ alloys are significantly influenced by the presence of two superconducting energy gaps \cite{shy14}. To investigate the properties of surface superconductivity, the $\rho(T)$ and $C(T)$ curves were plotted in the same temperature window in zero field [figures \ref{fig:surfacesuperconductivity} $(a)$ and $(c)$] as well as in a higher magnetic field ($> H_{C1}$) [figures \ref{fig:surfacesuperconductivity} $(b)$ and $(d)$]. In all the panels of Fig. \ref{fig:surfacesuperconductivity}, the (blue) solid circles (dots) represent the $C(T)$ measured in different $H$ (specified in each of the panels), the (black) solid triangles represent the $\rho(T)$ measured before coating the samples with Cu or Ag, and the (red) open squares represent the $\rho(T)$ measured after putting these metal coatings. Fig. \ref{fig:surfacesuperconductivity} (a) and (c) show that in zero field, the signature of the superconducting transition in the $\rho(T)$ and $C(T)$ curves are observed at the same temperature approximately, and this signature in the $\rho(T)$ curves is not affected by the presence or absence of the metal coatings on the samples. In the presence of applied magnetic field greater than $H_{c1}$ (figures \ref{fig:surfacesuperconductivity} (b) and (d)), the signature of this phase transition in the not-coated samples are observed at a higher temperature in the $\rho(T)$ as compared to that in the $C(T)$. However, after applying the metal coatings, in the presence of the same $H$, the signature of this phase transition in the $\rho(T)$ is observed to move towards lower temperatures and closer to that observed in the $C(T)$. The (four) probes used for the $\rho(T)$ measurement are on the sample surface, hence, this measurement is influenced by the surface related phenomenon. The suppression of the normal to superconducting transition temperature observed in $\rho(T)$ by applying a metal coating strongly suggests that this transition in the presence of magnetic field is indeed due to surface superconductivity \cite{GY64, FI66}. The $C(T)$ measurement, on the other hand, is influenced mainly by the bulk phenomenon. Figure \ref{fig:surfacesuperconductivity} thus indicates that the superconducting transitions corresponding to the bulk and surface are distinctly different in the present alloys. 

In this investigation, the focus is on the superconducting transition and its broadening observed with the help of the $\rho(T)$ measurements in the presence of applied magnetic fields. The present results (figures \ref{fig:RTMS}, \ref{fig:surfacesuperconductivity}), in conjugation with the flux-spot model (Fig. \ref{fig:fluxspot}) \cite{HP-1967}, suggest that the nature of mixed state formed within the surface sheath might give rise to the broadening of the superconducting transition in these alloys. This is supported by the analysis of $\rho(T)$ curves in different fields, presented below. 

In zero magnetic field the superconducting transition is quite sharp ($\Delta T_c < 0.1$ K) for both the Mo$_{1-x}$Re$_x$ alloys. However, the superconducting transition is increasingly broadened with the increase of applied magnetic field. Additionally, the superconducting transition in the presence of magnetic field exhibits a rounding-off behaviour near the onset of the superconducting transition and a tailing effect near the completion. The rounding-off behaviour of $\rho(T)$ is possibly related to the superconducting fluctuations \cite{R-1969}. The tail region of the superconducting transition becomes more significant as we go to higher magnetic fields, with a gradual suppression of the temperature corresponding to zero resistivity. Similar features have been observed in many bulk systems such as: Iron based superconductors: SmFeAsO$_{0.85}$ \cite{HMJH-2010}, BaFe$_2$As$_2$ \cite{SXMSKC-2012}, the high-$T_c$ superconductors \cite{TZHJYLS-2003, JM-1995} and even in some of the low $T_c$ superconductors: Nb thin films and Nb/Cu superlattices \cite{JJ-2005, vil05}, narrow strip of Nb \cite{yoi93}, In film \cite{oku97}, amorphous Mo$_x$Si$_{1-x}$ and Mo$_3$Si films \cite{oku01, yeh93}. Both the tailing-off behaviour as well as the broadening of the superconducting transition are generally attributed to the effect of thermal fluctuations in  the superconductors, and it gives rise to a rich variety of flux-line dynamics in the presence of quenched disorder and other defects \cite{DMD-1991}. Generally, the effect of thermal fluctuations is expected to be small in the case of the low $T_c$ superconductors. However, the effect of thermal fluctuations is greatly enhanced in the case of the two-dimensional (2d) superconductor as compared to the three-dimensional (3d) ones \cite{DMD-1991}. 

It is known that, these alloys are unlikely to be 2d superconductors \cite{shy15b}. On the other hand, the Ginzburg number, which quantifies the effect of thermal fluctuations in the superconductors, is found to be $5.45 \times 10^{-9}$ and $1.04 \times 10^{-8}$ for the $x$ = 0.20 and 0.25 Mo$_{1-x}$Re$_x$ alloys respectively. The Ginzburg number for Mo$_{1-x}$Re$_x$ alloys were estimated using the following relation \cite{sok93}.

\begin{equation} \label{eq:ginzburgnumber}
G_i = 32 {\pi}^4 \left(\frac{k_B T_c \kappa \lambda_{GL}(0)}{\Phi_0^2}\right)^2
\end{equation}
where, $k_B$ is the Boltzman constant, $T_c$ is the superconducting transition temperature, $\kappa$ is the Ginzburg-Landau parameter, $\lambda_{GL}(0)$ is the Ginzburg-Landau penetration depth at absolute zero temperature and $\Phi_0$ is the flux quantum. To estimate the value of Ginzburg number using Eq. \ref{eq:ginzburgnumber}, we have used $\kappa = \frac{H_{c2}(0)}{\sqrt{2}H_c(0)}$. The values of $H_{c2}(0)$ (the upper critical field at absolute zero temperature) have been taken from ref.\cite{shy15b}. The thermodynamic critical field, $H_c(0)$, and $\lambda_{GL}(0)$ are estimated from the heat capacity measurements and using $\lambda_{GL}(0) = \kappa \times \xi_{GL}(0)$, where $\xi_{GL}(0) = \left(\frac{\Phi_0}{2\pi H_{c2}}\right)^{\frac{1}{2}}$ respectively. In this context, it may be noted that while the Ginzburg number generally comes out to be $10^{-2}$-$10^{-3}$ for the high $T_c$ superconductors \cite{sok93, sok91, GMYAV-1994}, for the low $T_c$ superconductors this number is found to be in the range of $10^{-6}$-$10^{-8}$ \cite{GMYAV-1994}. Thus, the Ginzburg number for the Mo$_{1-x}$Re$_x$ alloys is similar to those of the conventional low $T_c$ superconductors, indicating that the effect of thermal fluctuations may not be substantial in these alloys. In view of the experimentally observed signature of surface superconductivity in the present Mo$_{1-x}$Re$_x$ alloys, we believe that the formation of 2d pancake like flux-lines (Kulik vortex-state \cite{KUL-1969, KUL-1967}) within the surface sheath causes an enhancement of the effect of the thermal fluctuations in these alloys. Therefore, similar to other systems \cite{SGXJH-2005, AMYMKN-2001}, the analysis of the $\rho(T)$ behaviour is performed using the thermally activated flux-flow (TAFF) model \cite{pal88}.

\begin{figure}[h]
	\centering
	\includegraphics[height=11cm]{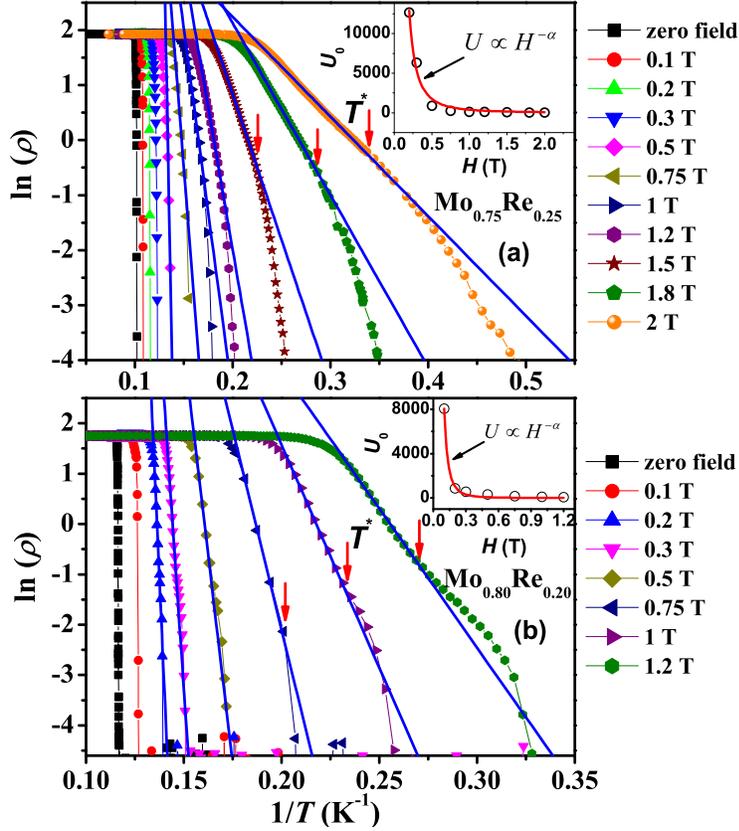}
	\caption{\label{fig:arrhenius} Arrhenius plots for $\rho(T)$ in different magnetic fields for the Mo$_{1-x}$Re$_x$ alloys. In both the panels ($a$ and $b$), the blue lines show the straight line fit to the experimental data in the TAFF region and the arrows indicate the $T^{\ast}$ value, where the experimental data deviates from linearity. In the TAFF region, the activation energy $U_0(H)$ is estimated, which is shown in the insets to both the panels for the respective alloys. The $U_0(H)$ shows a power law behaviour in the TAFF region, with $\alpha \geq 2$ for both the alloys.}
	\end{figure}

In the TAFF model, the temperature dependence of electrical resistivity is described by the Arrhenius relation \cite{pal88},
\begin{equation} \label{eq:8-2}
\rho(T) = \rho_0 \exp\left[-U(H,T)/T\right]
\end{equation}
where, $\rho_0$ is the pre-exponential factor independent of field and $U(H,T)$ is the activation energy. Arrhenius relation suggests that the $ln \rho$ vs. $1/T$ should be linear in the TAFF region.

Figure \ref{fig:arrhenius} shows the Arrhenius plots for $\rho(T)$ in different magnetic fields for the Mo$_{1-x}$Re$_x$ ($x$ = 0.20, 0.25) alloys. The activation energy ($U_0$) is estimated by fitting a straight line (blue) to the experimental data in the TAFF region. The fitted straight line (blue) in both the panels of Fig. \ref{fig:arrhenius}, shows a deviation from linearity at temperature $T^{\ast}$. In Fig. \ref{fig:arrhenius}, the insets to the panels $a$ and $b$ show the activation energy $U_0(H)$. For both the alloys, the $U_0(H)$ shows a power law behaviour ($U = A H^{-\alpha}$) with $\alpha \geq 2$. High values of $\alpha$ has also been reported in other superconductors such as, iron pnictide and cuprates and has been described in terms of the collective flux-line pinning behaviour \cite{lei10, ge14, TZHJYLS-2003}. 

	\begin{figure}[h]
	\centering
	\includegraphics[height=11cm]{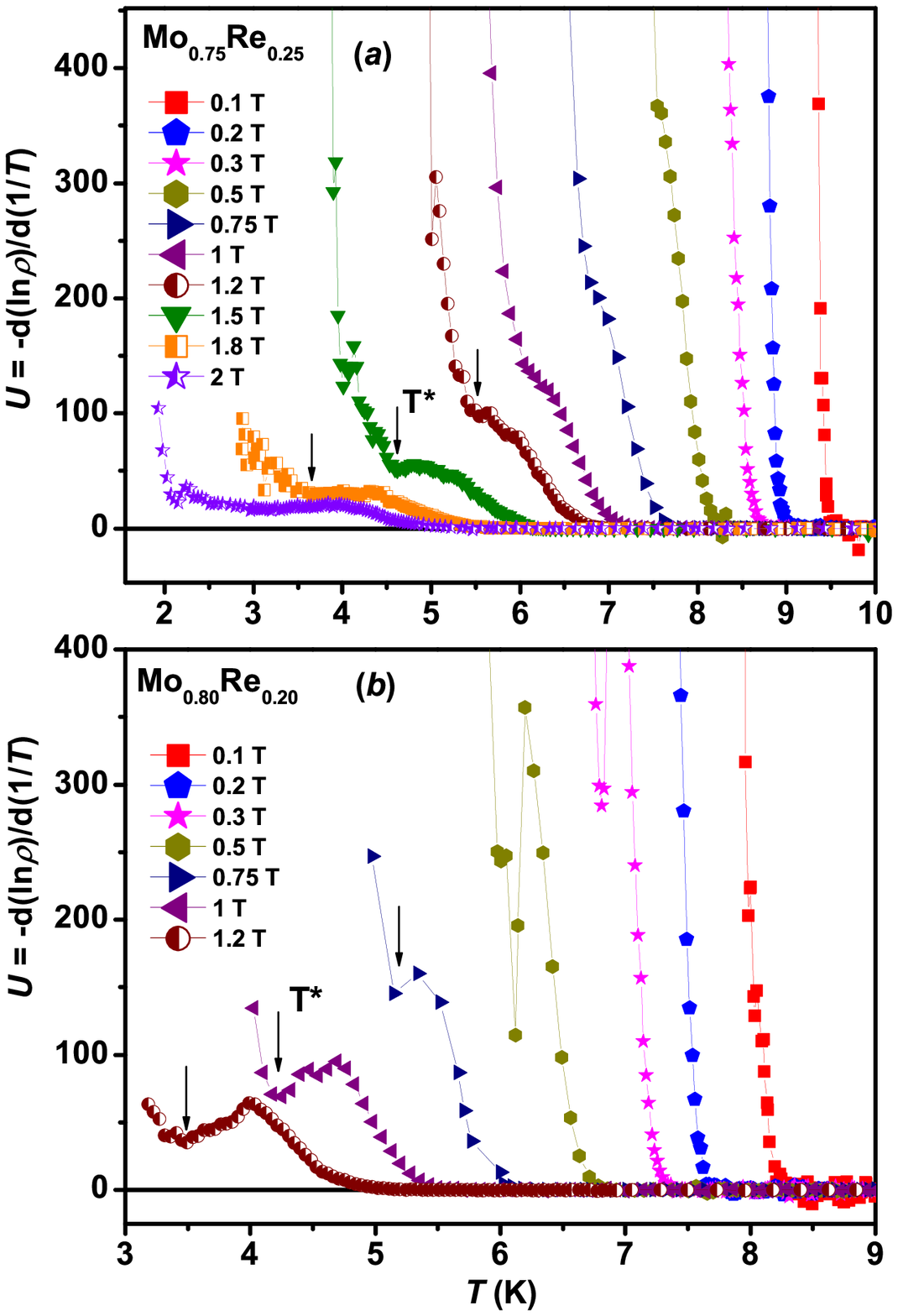}
	\caption{\label{fig:activation} Temperature dependence of activation energy, $U = -d(\ln\rho)/d(1/T)$, in the Mo$_{1-x}$Re$_x$ alloys. In both the panels, for each curve, the arrow indicates the $T^{\ast}$ value, which is the temperature below which the 2d flux-lines enter into the vortex-liquid to vortex-glass critical transformation region.}
	\end{figure}
	
To analyze the experimental data below $T^{\ast}$, we have estimated the activation energy, $U$(T), using Eq. \ref{eq:8-2}, 

\begin{equation} \label{eq:8-3}
U = -d(\ln\rho)/d(1/T)
\end{equation}

\begin{figure}[h]
	\centering
	\includegraphics[height=11cm]{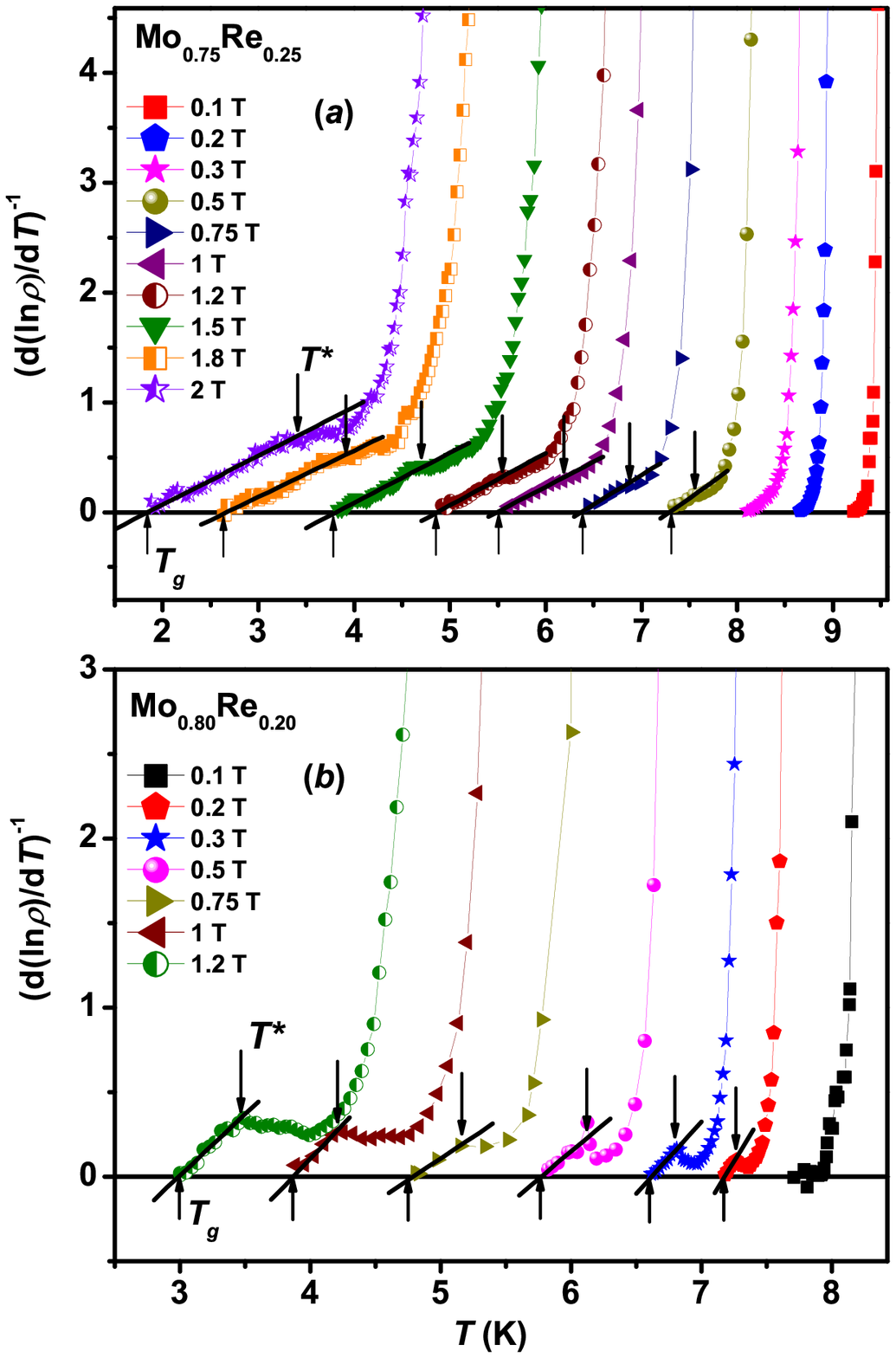}
	\caption{\label{fig:inverse} Temperature dependence of inverse logarithmic derivative of resistivity in the Mo$_{1-x}$Re$_x$ alloys in different magnetic fields. In both the panels, for each curve, the solid lines are fit to the experimental data and show the linear behaviour in the tail region.}
	\end{figure}

The temperature dependence of the activation energy of the Mo$_{1-x}$Re$_x$ alloys are shown in Fig. \ref{fig:activation}. The activation energy increases rapidly with decreasing temperature below a characteristic temperature $T^{\ast}$, which matches exactly with that pointed out in Fig. \ref{fig:arrhenius} as the temperature corresponding to the deviation from linearity. The rapid increase of activation energy below $T^{\ast}$ marks the entry into a critical regime associated with the vortex-liquid to vortex-glass transformation, as previously observed in the other superconductors \cite{TZHJYLS-2003, HMJH-2010, HPD-1992}. Figure \ref{fig:inverse} shows the temperature dependence of inverse logarithmic derivative of resistivity with the values of $T^{\ast}$ marked by arrows. Thus, to investigate into the existence of vortex-glass state, we need to study the temperature dependence of resistivity below the characteristic temperature $T^{\ast}$.

According to the vortex glass theory \cite{DMD-1991}, the electrical resistivity vanishes at the glass transition temperature $T_g$,

\begin{equation} \label{eq:8-4}
\rho \propto (T - T_g)^s
\end{equation}
Where, $s$ is the critical exponent, defined as, $s = \nu(z+2-d)$, here $\nu$ is the static exponent, $z$ is the dynamic exponent and $d$ is a dimensional factor ($d =2$ in the case of surface mixed state). The glass temperature $T_g$ is estimated by applying Eq. \ref{eq:8-5} to the tail region of the $\rho(T)$ curves.

\begin{equation} \label{eq:8-5}
\left(\frac{d(\ln\rho)}{dT}\right)^{-1} = \frac{1}{s} \left(T - T_g\right)
\end{equation}

According to Eq. \ref{eq:8-5}, the inverse logarithmic derivative of resistivity is linearly proportional to the temperature of measurement. Fig. \ref{fig:inverse} shows that the resistivity in the temperature range $T_g < T < T^{\ast}$ is well described by Eq. \ref{eq:8-5}. In Fig. \ref{fig:inverse}, the slope of the straight line fitted to the experimental data (within the temperature range $T_g < T < T^{\ast}$), gives the value of the critical exponent and its intercept on the temperature axis gives the value of glass transition temperature $T_g$. In figure \ref{fig:inverse}, the deviation from linearity at temperature $T^{\ast}$, marks the upper limit of the critical-region associated with the vortex-glass to vortex-liquid phase transformation. Fig. \ref{fig:activation} and \ref{fig:inverse} show that the $\rho (T)$ in the temperature range, $T_g < T < T^{\ast}$, is well described by the vortex-glass model. 

\begin{figure}[h]
	\centering
	\includegraphics[height=6cm]{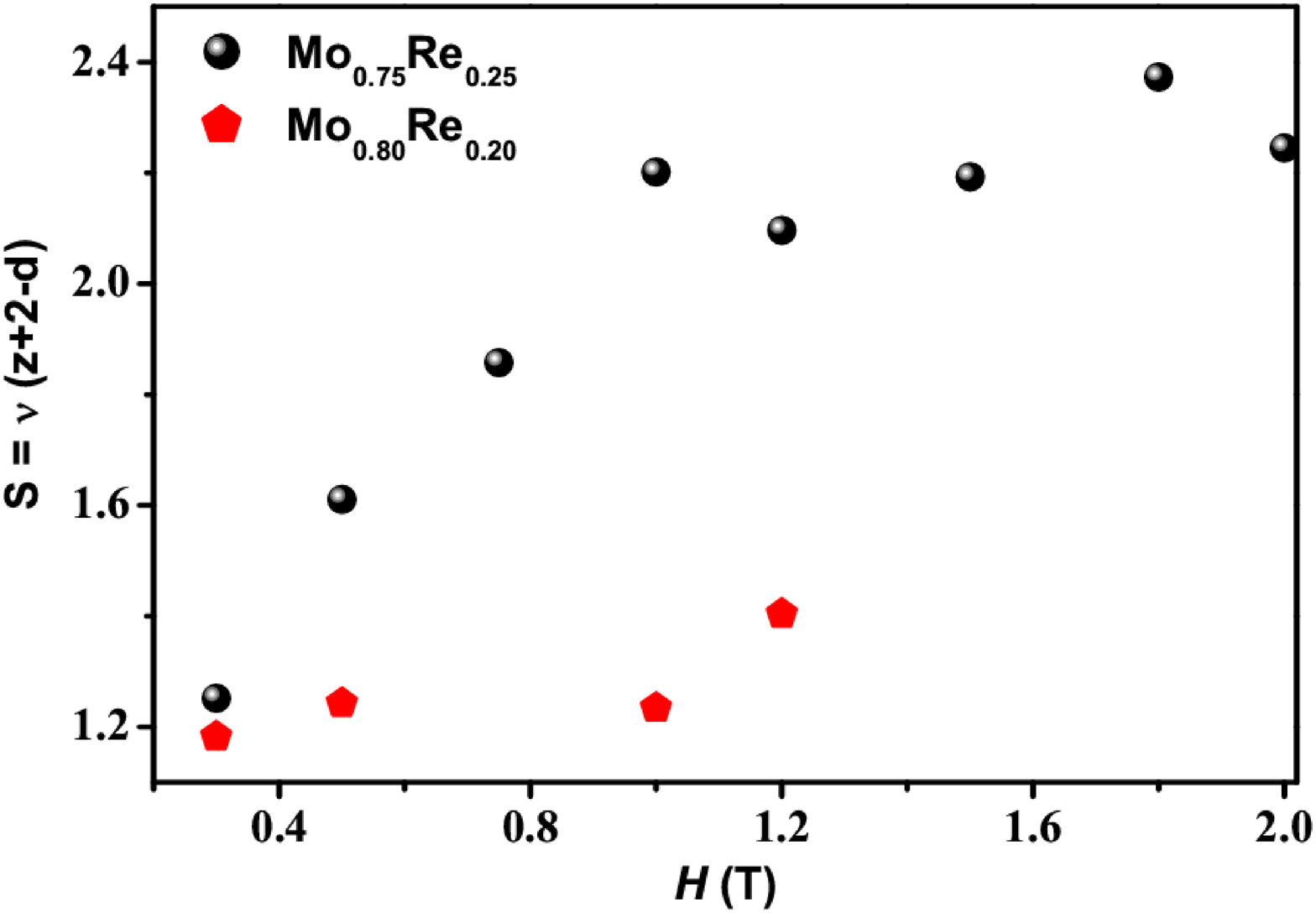}
	\caption{\label{fig:exponent} Critical exponent $s = \nu (z+2-d)$, in different magnetic fields for the Mo$_{1-x}$Re$_x$ alloys. The small value (less than 2.7 for both alloys) of the critical exponent is attributed to the 2d nature of the vortex-glass state \protect\cite{HMJH-2010}.}
	\end{figure}

The magnetic field dependence of the critical exponents is shown in Fig. \ref{fig:exponent}. The critical exponents increase with increasing magnetic field for both the alloys. As seen in Fig. \ref{fig:exponent}, the maximum attainable values of the critical exponents for the $x$ = 0.20 and $x$ = 0.25 alloys are about 1.4 at $H$~=~1.2 T and 2.4 at $H$~=~1.8 T respectively. According to the vortex-glass model \cite{GMYAV-1994}, the critical exponent values are in between 2.7 - 8.5 for 3d vortex-glass state \cite{HMJH-2010}. On the other hand, the maximum permissible value of critical exponent for the 2d vortex-glass state is reported to be 2.7 \cite{HMJH-2010}. The small values of critical exponent in the present alloys indicate that the vortex-glass state has formed in the surface sheath (Kulik-vortex state). 

It may be noted that in literature the small value of critical exponent has also been attributed to the Bose-glass phase. The Bose-glass phase is formed due to the interaction of the flux-lines with correlated disorder, such as the twin boundaries and the columnar defects \cite{GMYAV-1994, SEECGF-1998, RWLADG-2002}. However, optical metallography studies and the published literature  on these alloys \cite{shy15b}, do not show any indication of the presence of correlated disorder (twin boundaries, columnar defects). 

	\begin{figure}[h]
	\centering
	\includegraphics[height=10cm]{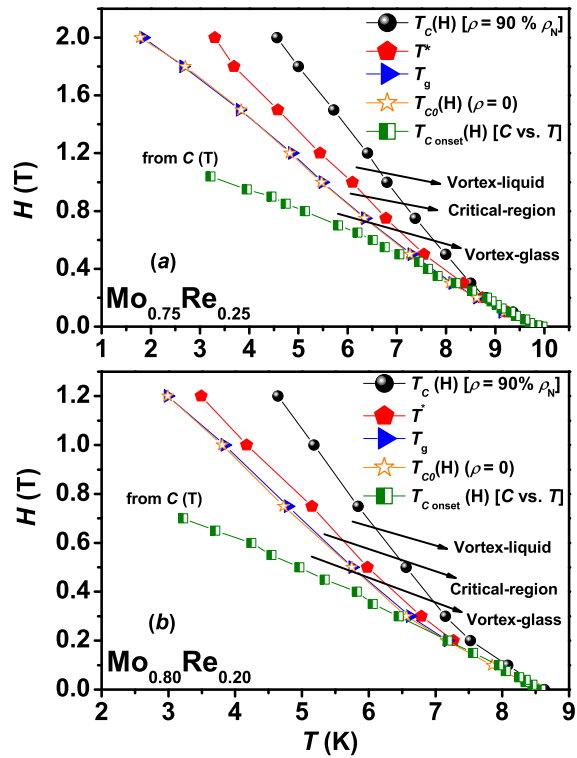}
	\caption{\label{fig:HTdiagram} The $H$-$T$ phase-diagrams for the Mo$_{1-x}$Re$_x$ superconductors.}
	\end{figure}

Figure \ref{fig:HTdiagram} shows the detailed $H$-$T$ phase diagrams for both the Mo$_{1-x}$Re$_x$ alloys. The characteristic values in the phase diagrams are obtained through $\rho(T)$ and $C(T)$ measurements in different magnetic fields. In Fig. \ref{fig:HTdiagram}, the uppermost line with solid spheres shows the $T_c(H)$ values, estimated from $\rho(T)$ in different applied magnetic fields. The $T_c(H)$ values were estimated, as the temperature corresponding to the 90\% of the normal state resistivity $(\rho_N)$. Above this line, both the alloys are in the normal state. The line with solid diamonds represents the $T^{\ast}$ line, which corresponds to the upper limit of the vortex-glass critical-regime. The region between the $T_c(H)$ line and the $T^{\ast}$ line corresponds to the vortex-liquid behaviour, where the system shows non-zero resistance because of flux flow. The region between the $T^{\ast}$ and $T_g$ lines is demarcated as the vortex-liquid to vortex-glass critical transition region. The $T_g$ line corresponds to the glass temperature below which the surface flux-lines freeze to form the vortex-glass state within the surface sheath i.e., the Kulik vortex-state. The $T_{c0}$ line (open stars) estimated from the $\rho(T)$ curves marks the temperature, where resistivity goes to zero in the respective field. It is clearly observed in Fig. \ref{fig:HTdiagram} that this zero resistivity line matches exactly with the $T_g$ line which further confirms the existence of the vortex-glass state in Mo$_{1-x}$Re$_x$ superconductors. The onset of bulk superconducting transition were estimated from the $C(T)$ measurements in different magnetic fields. The $T_{c_{onset}}(H)$ values for bulk superconducting transition is defined as the temperature, where the superconducting transition just starts (onset of the peak in $C(T)$). We observe in Fig. \ref{fig:HTdiagram} that all aspects related to the vortex-liquid to vortex-glass transformation takes place above the bulk superconducting transition temperature. This phase diagram thus provides additional evidence of the formation of the Kulik vortex-state. 

\section{Summary and Conclusion}

In summary, the temperature dependence of electrical resistivity and heat capacity were measured experimentally, and, the signature of vortex-liquid to vortex-glass transformation within the surface mixed state or Kulik vortex-state in the Mo$_{1-x}$Re$_x$~($x$ = 0.20, 0.25) alloys were observed in these measurements. The vortex-glass critical-regime were investigated using the vortex-glass theory \cite{DMD-1991}. The critical exponent, estimated from the vortex-glass theory increases with the increase in magnetic field, and nearly saturates at small critical exponent values (less than 2.7) for both the alloys. In the absence of correlated disorder, such small critical exponents indicate the possible existence of 2d vortex-glass state. Different characteristic temperatures, associated with the vortex-glass transformation were estimated through analysis and the detailed $H$-$T$ phase diagrams were constructed. The $H$-$T$ phase diagrams clearly show that the phase transformations associated with vortex-liquid to vortex-glass transition are taking place in the 2d surface mixed state, above the bulk superconducting transition  temperature corresponding to the particular magnetic field applied.

\section*{Acknowledgements}
Authors would like to thank to Prof. Lesley F Cohen, Prof. Luis Ghivelder and Prof. Said Salem Sugui Jr. for interesting discussions and important suggestions. 

\section*{References}

%\bibliographystyle{iopart-num}
%\bibliography{VG}

\begin{thebibliography}{10}
\expandafter\ifx\csname url\endcsname\relax
  \def\url#1{{\tt #1}}\fi
\expandafter\ifx\csname urlprefix\endcsname\relax\def\urlprefix{URL }\fi
\providecommand{\eprint}[2][]{\url{#2}}
% Bibliography created with iopart-num v2.1
% /biblio/bibtex/contrib/iopart-num

\bibitem{DP-1963}
Saint-James D and de~Gennes P~G 1963 {\em Phys. Lett\/} {\bf 7} 306

\bibitem{CY-1964}
Hempstead C~F and Kim Y~B 1964 {\em Phys. Rev. Lett.\/} {\bf 12} 145

\bibitem{GY64}
Gygax S and Kropschot R~H 1964 {\em Phys. Lett.\/} {\bf 9} 91

\bibitem{FI66}
Fischer G and Klein R 1966 {\em Phys. Lett.\/} {\bf 23} 311

\bibitem{Mic96}
Tinkham M 1996 {\em Introduction to Superconductivity\/} 2nd ed (Mineola, New
  York: Dover publications, Inc.)

\bibitem{J-1964}
Park J~G 1964 {\em Rev. Mod. Phys.\/} {\bf 36} 87

\bibitem{HL-1965}
Fink H~J and Barnes L~J 1965 {\em Phys. Rev. Lett.\/} {\bf 15} 792

\bibitem{LH-1966}
Barnes L~J and Fink H~J 1966 {\em Phys. Rev.\/} {\bf 149} 186

\bibitem{Das08}
Das P, Tomy C~V, Banerjee S~S, Takeya H, Ramakrishnan S and Grover A~K 2008
  {\em Phys. Rev. B\/} {\bf 78} 214504

\bibitem{kar70}
Karasik V~R and Shebalin I~Y 1970 {\em Sov. Phys. JETP\/} {\bf 30} 1068

\bibitem{men06}
Tsindlekht M~I, Leviev G~I, Genkin V~M, Felner I, Paderno Y~B and Filippov V~B
  2006 {\em Phys. Rev. B\/} {\bf 73} 104507

\bibitem{ryd03}
Rydh A, Welp U, Hiller J~M, Koshelev A~E, Kwok W~K, , Crabtree G~W, Kim K~H~P,
  Kim K~H, Jung C~U, Lee H~S, Kang B and Lee S~I 2003 {\em Phys. Rev. B\/} {\bf
  68} 172502

\bibitem{abr65}
Abrikosov A~A 1965 {\em Soviet Phys.- JETP\/} {\bf 20} 480

\bibitem{par65}
Park J~G 1965 {\em Phys. Rev. Lett.\/} {\bf 15} 352

\bibitem{par66}
Park J~G 1966 {\em Phys. Rev. Lett.\/} {\bf 16} 1196

\bibitem{fin65}
Fink H~J 1965 {\em Phys. Rev. Lett.\/} {\bf 14} 309

\bibitem{HP-1967}
H~R~Hart J and Swartz P~S 1967 {\em Phys. Rev.\/} {\bf 156} 403

\bibitem{MRW-1969}
Beasley M~R, Labusch R and Webb W~W 1969 {\em Phys. Rev.\/} {\bf 181} 682

\bibitem{JJJ-1970}
Gosselin J, Silcox J and Trefny J~U 1970 {\em Phys. Rev. B\/} {\bf 2} 4508

\bibitem{PBY-1993}
Mathieu P, Placais B and Simon Y 1993 {\em Phys. Rev. B\/} {\bf 48} 7376

\bibitem{AP-2004}
Pan A~V and Esquinazi P 2004 {\em Phys. Rev. B\/} {\bf 70} 184510

\bibitem{KUL-1969}
Kulik I~O 1969 {\em Sov. Phys. JETP\/} {\bf 28} 461

\bibitem{KUL-1967}
Kulik I~O 1967 {\em Sov. Phys. JETP\/} {\bf 25} 1085

\bibitem{MTA-1991}
Fisher M~P~A, Tokuyasu T~A and Young A~P 1991 {\em Phys. Rev. Lett.\/} {\bf 66}
  2931

\bibitem{DMD-1991}
Fisher D~S, Fisher M~P~A and Huse D~A 1991 {\em Phys. Rev. B\/} {\bf 43} 130

\bibitem{CPRBA-1992}
Dekker C, Woltgens P~J~M, Koch R~H, Hussey B~W and Gupta A 1992 {\em Phys. Rev.
  Lett.\/} {\bf 69} 2717

\bibitem{MS-1999}
Choi M~Y and Park S~Y 1999 {\em Phys. Rev. B.\/} {\bf 60} 4070

\bibitem{B-2000}
Kim B~J 2000 {\em Phys. Rev. B\/} {\bf 62} 644

\bibitem{Y-1992}
Li Y~H 1992 {\em Phys. Rev. Lett.\/} {\bf 69} 1819

\bibitem{PBP-2003}
Holme P, Kim B~J and Minnhagen P 2003 {\em Phys. Rev. B\/} {\bf 67} 104510

\bibitem{shy14}
Sundar S, Chandra L~S~S, Chattopadhyay M~K and Roy S~B 2015 {\em J. Phys.:
  Condens. Matter\/} {\bf 27} 045701

\bibitem{shy15a}
Sundar S, Chandra L~S~S, Chattopadhyay M~K, Pandey S~K, Venkateshwarlu D, Rawat
  R, Ganesan V and Roy S~B 2015 {\em New J. Phys.\/} {\bf 17} 053003

\bibitem{and89}
Andreone A, Barone A, Chiara A~D, Fontana F, Mascolo G, Palmieri V, Peluso G,
  Pepe G and Scotti D~U~U 1989 {\em J. Supercond.\/} {\bf 2} 493

\bibitem{LER-1966}
Lerner E and Daunt J~G 1966 {\em Phys Rev\/} {\bf 142} 251

\bibitem{JOI-64}
Joiner W~C~H and Blaugher R~D 1964 {\em Rev. Mod. Phys.\/} {\bf 36} 67

\bibitem{shy15b}
Sundar S, Chattopadhyay M~K, Chandra L~S~S and Roy S~B 2015 {\em Physica C\/}
  {\bf 519} 13

\bibitem{MOR-1963}
Morin F~J and Maita J~P 1963 {\em Phys Rev\/} {\bf 129} 1115

\bibitem{CRXS-2010}
Wang C~C, Zeng R, Xu X and Dou S~X 2010 {\em J. Appl. Phys.\/} {\bf 108} 093907

\bibitem{JSCM-1998}
Mosqueira J, Curras S~R, Carballeira C, Ramallo M~V, Siebold T, Torron C,
  ACampa J, Rasines I and Vidal F 1998 {\em Supercond. Sci. Technol.\/} {\bf
  11} 821

\bibitem{AMZYKHHT-2002}
Pradhan A~K, Tokunaga M, Shi Z~X, Takano Y, Togano K, Kito H, Ihara H and
  Tamegai T 2002 {\em Phys. Rev. B\/} {\bf 65} 144513

\bibitem{Yu-2003}
Eltsev Y 2003 {\em Physica C\/} {\bf 385} 162

\bibitem{HYYSIGK-2009}
Kim H~J, Liu Y, Oh Y~S, Khim S, Kim I, Stewart G~R and Kim K~H 2009 {\em Phys.
  Rev. B\/} {\bf 79} 014514

\bibitem{BAYOM-1998}
Lundqvist B, Rydh A, Eltsev Y, Rapp and Andersson M 1998 {\em Phys. Rev. B\/}
  {\bf 57} R14064

\bibitem{UAGWG-2003}
Welp U, Rydh A, Karapetrov G, Kwok W~K, Crabtree G~W, Marcenat C, Paulius L,
  Klein T, Marcus J, Kim K~H~P, Jung C~U, Lee H~S, Kang B and Lee S~I 2003 {\em
  Phys. Rev. B\/} {\bf 67} 012505

\bibitem{AJSJH-2002}
Sologubenko A~V, Jun J, Kazakov S~M, Karpinski J and Ott H~R 2002 {\em Phys.
  Rev. B\/} {\bf 65} 180505(R)

\bibitem{R-1969}
Hake R~R 1969 {\em Phys. Rev. Lett.\/} {\bf 23} 1105

\bibitem{HMJH-2010}
Lee H~S, Bartkowiak M, Kim J~S,  and Lee H~J 2010 {\em Phy. Rev. B\/} {\bf 82}
  104523

\bibitem{SXMSKC-2012}
Ghorbani S~R, Wang X~L, Shabazi M, Dou S~X, Choi K~Y and Lin C~T 2012 {\em
  Appl. Phys. Lett.\/} {\bf 100} 072603

\bibitem{TZHJYLS-2003}
Yang T, Wang Z~H, Zhang H, J~Fang Y~N, Qiu L and Ding S~Y 2003 {\em Physica
  C\/} {\bf 384} 130

\bibitem{JM-1995}
Deak J and McElfresh M 1995 {\em Phys. Rev. B\/} {\bf 52} R3880

\bibitem{JJ-2005}
Villegas J and Vicent J~L 2005 {\em Phys. Rev. B\/} {\bf 71} 144522

\bibitem{vil05}
Villegas J~E, Gonzalez E~M, Sefrioui Z, Santamaria J and Vicent J~L 2005 {\em
  Phys. Rev. B\/} {\bf 72} 174512

\bibitem{yoi93}
Ando Y, Kubota H and Tanaka S 1993 {\em Phys. Rev. B\/} {\bf 48} 7716

\bibitem{oku97}
Okuma S and Kokubo N 1997 {\em Phys. Rev. B\/} {\bf 56} 14138

\bibitem{oku01}
Okuma S, Imamoto Y and Morita M 2001 {\em Phys. Rev. Lett.\/} {\bf 86} 3136

\bibitem{yeh93}
Yeh N~C, Reed D~S, Jiang W, Kriplani U, Tsuei C~C, Chi C~C and Holtzberg F 1993
  {\em Phys. Rev. Lett.\/} {\bf 71} 4043

\bibitem{sok93}
Sokolov A, Kufaev Y~A and Sonin E 1993 {\em Physica C\/} {\bf 212} 19

\bibitem{sok91}
Sokolov A 1991 {\em Physica C\/} {\bf 174} 208

\bibitem{GMYAV-1994}
Blatter G, Feigelman M~Y, BGeshkenbein Y, Larkin A~I and Vinokur V~M 1994 {\em
  Rev. Mod. Phys.\/} {\bf 66} 1125

\bibitem{SGXJH-2005}
Liu S~L, Wu G~J, Xu X~B, Wu J and Shao H~M 2005 {\em Supercond. Sci.
  Technol.\/} {\bf 18} 1332

\bibitem{AMYMKN-2001}
Pradhan A~K, Muralidhar M, Feng Y, Murakami M, Nakao K and Koshizuka N 2001
  {\em Phys. Rev. B\/} {\bf 64} 172505

\bibitem{pal88}
Palstra T~T~M, Batlogg B, Schneemeyer L~F and Waszczak J~V 1988 {\em Phys. Rev.
  Lett.\/} {\bf 61} 1662

\bibitem{lei10}
Lei H, Hu R, Choi E~S and Petrovic C 2010 {\em Phys. Rev. B\/} {\bf 82} 134525

\bibitem{ge14}
Ge J, Gutierrez J, Li J, Yuan J, Wang H~B, Yamaura K, Takayama-Muromachi E and
  Moshchalkov V~V 2014 {\em Appl. Phys. Lett\/} {\bf 104} 112603

\bibitem{HPD-1992}
Safar H, Gammel P~L and Bishop D~J 1992 {\em Phys Rev. Lett\/} {\bf 68} 2672

\bibitem{SEECGF-1998}
Grigera S~A, Morre E, Osquiguil E, Balseiro C, Nieva G and de~la Cruz F 1998
  {\em Phys. Rev. Lett\/} {\bf 81} 2348

\bibitem{RWLADG-2002}
Olsson R~J, Kwok W~K, Paulius L~M, Petrean A~M, Hofman D~J and Crabtree G~W
  2002 {\em Phys. Rev. B\/} {\bf 65} 104520

\end{thebibliography}
\providecommand{\newblock}{}

\end{document}